\documentclass{sigchi}

\CopyrightYear{2019}
\setcopyright{acmlicensed}

\usepackage{balance}       
\usepackage{graphics}      
\usepackage[T1]{fontenc}   
\usepackage{txfonts}
\usepackage{mathptmx}
\usepackage[pdflang={en-US},pdftex]{hyperref}
\usepackage{color}
\usepackage{booktabs}
\usepackage{textcomp}
\usepackage{etoolbox}

\usepackage{microtype}        
\usepackage{ccicons}          

\usepackage{todonotes}

\def\plaintitle{Safe Walking In VR using Augmented Virtuality}

\def\emptyauthor{}
\def\plainkeywords{Virtual locomotion interfaces, Confined and cluttered spaces, Natural Walking, Walking in Place, Affordable Virtual Reality, Augmented Virtual Reality}

\makeatletter
\def\url@leostyle{%
  \@ifundefined{selectfont}{
    \def\UrlFont{\sf}
  }{
    \def\UrlFont{\small\bf\ttfamily}
  }}
\makeatother
\def\pprw{8.5in}
\def\pprh{11in}

\definecolor{linkColor}{RGB}{6,125,233}
\hypersetup{%
  pdftitle={\plaintitle},
  pdfauthor={\emptyauthor},
  pdfkeywords={\plainkeywords},
  pdfdisplaydoctitle=true, 
  bookmarksnumbered,
  pdfstartview={FitH},
  colorlinks,
  citecolor=black,
  filecolor=black,
  linkcolor=black,
  urlcolor=linkColor,
  breaklinks=true,
  hypertexnames=false
}

\makeatletter
\def\@copyrightspace{\relax}
\makeatother

\begin{document}

\title{\plaintitle}

\author{%
  \alignauthor{Maur\'{i}cio Sousa, Daniel Mendes, and Joaquim Jorge\\
    \affaddr{INESC-ID Lisboa, Instituto Superior T\'{e}cnico, Universidade de Lisboa}\\
    \email{\{antonio.sousa, danielmendes, jorgej\}@tecnico.ulisboa.pt}}
    }

\maketitle

\begin{abstract}
    New technologies allow ordinary people to access Virtual Reality at affordable prices in their homes.
One of the most important tasks when interacting with immersive Virtual Reality is to navigate the virtual environments (VEs). Arguably, the best methods to accomplish this use direct control interfaces. Among those, natural walking (NW) makes for an enjoyable user experience.
However, common techniques to support direct control interfaces in VEs feature constraints that make it difficult to use those methods in cramped home environments. Indeed, \textit{NW} requires unobstructed and open space, to allow users to roam around without fear of stumbling on obstacles while immersed in a virtual world.

To approach this problem, we propose a new  virtual locomotion technique, which we call \textit{Combined Walking in Place} (CWIP). 
CWIP allows people to take advantage of the available physical space and empowers them to use \textit{NW} to navigate in the virtual world. For longer distances, we adopt \textit{Walking in Place} (WIP) to enable them to move in the virtual world beyond the confines of a cramped real room.

However, roaming in an immersive alternate reality, while moving in the confines of a cluttered environment can lead people to stumble and fall. To approach these problems we developed a technique called \textit{Augmented Virtual Reality} (AVR), to inform users about real world hazards, such as chairs, drawers, walls via \textit{proxies} and signs placed in the virtual world. 
We propose thus \textit{Combined Walking in Place in Augmented Virtual Reality} (CWIP-AVR) as a way to safely explore VR in the cramped confines of your own home. To our knowledge, this is the first approach to combined different locomotion modalities in a safe manner. We assessed its effectiveness in a user study with  20 participants to validate their ability to navigate a virtual world while walking in a confined and cluttered real space. Our results show that  CWIP-AVR allows people to navigate VR safely, while switching between locomotion modes flexibly and maintaining a good degree of immersion.

\end{abstract}



\section{Introduction}

New devices powered by the latest surge in VR ventures such as the \textit{Samsung GearVR}, \textit{Oculus Rift} and \textit{Steam VR} to name a few, allow ordinary users to access and enjoy affordable Virtual Reality in their homes.
Prior to these recent advances, only a select few working in well-funded research laboratories and private companies could afford such experiences.

Navigation is one of the most common and important tasks that people perform while interacting with VEs. 
This allows users to control their position when roaming VEs. Even when Navigation is not the main focus of the experience in a VE, it is surely an essential support task to the main goal~\cite{BowmanCompleto2004}.

Among the many possible ways to accomplish locomotion in VR, the technique that best conveys the sense of presence in the virtual world is \textit{Natural Walking}(NW).
\textit{Walking in Place}(WIP) is the best alternative~\cite{Usoh1999}.
However, current approaches to walking in VR entail requirements that preclude their adoption in  common settings. Chief among these, is need for large unencumbered areas that can be dedicated exclusively to  navigating the virtual world. Indeed, the current technology fosters and encourages immersion in cramped and confined environments such as living rooms, dorms and other home settings. NW while immersed becomes a recipe for disaster, as people become blissfully unaware of their surroundings. While gamer-like contraptions such as joysticks and driving seats, obviate the problem, they change the experience in unnatural ways, especially when exploring the VE calls for natural locomotion, instead of vehicular displacement.

\begin{figure}[t]
   \centering
    \includegraphics[width=\columnwidth]{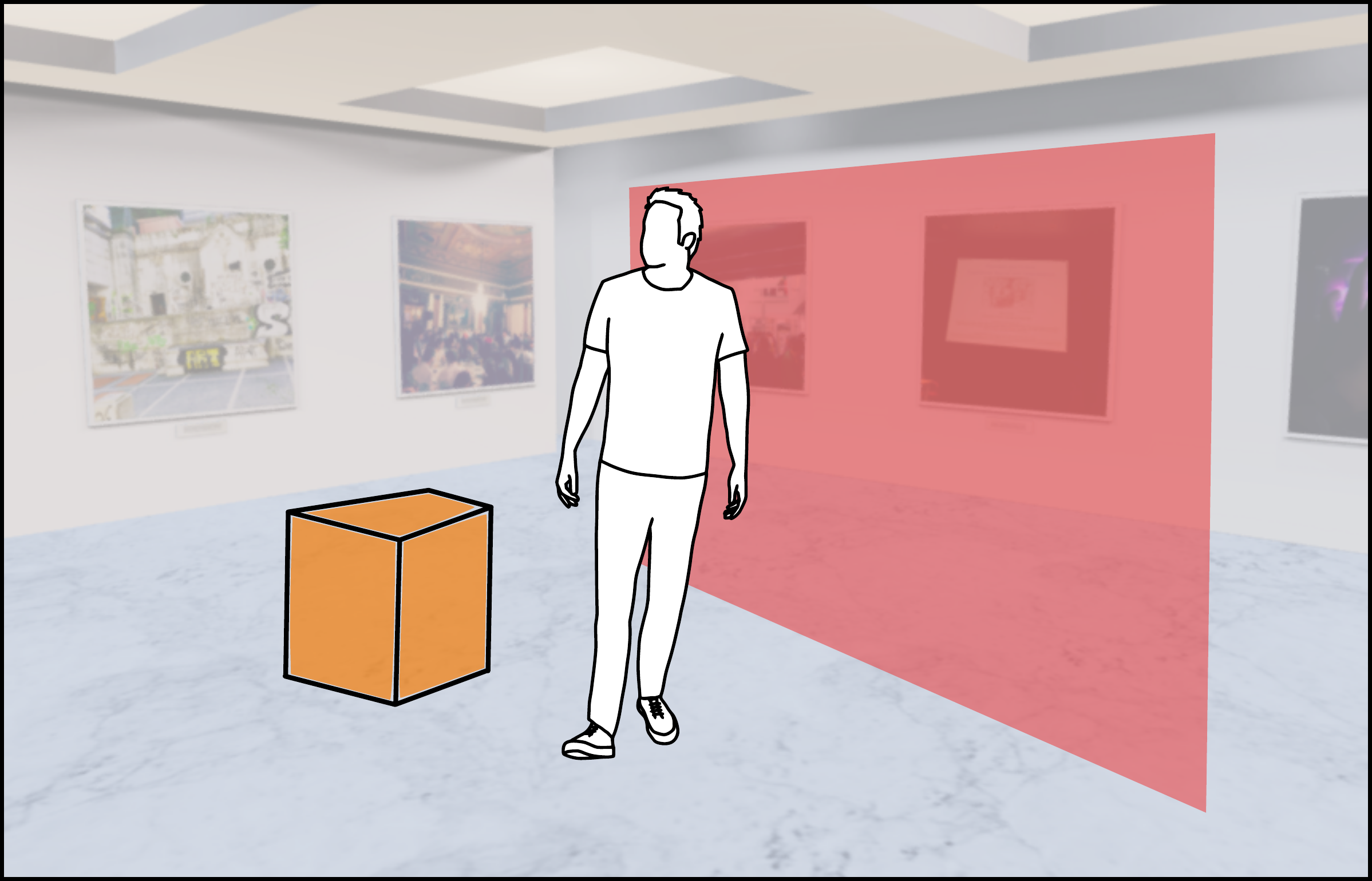}
    \caption{To achieve safe walking in VR, our approach combines  Walking-In-Place locomotion technique with Natural Walking and virtual proximity indicators for obstacles and room boundaries.}
	\label{fig:teaser}
\end{figure}

To approach this issue, we propose a new navigation method, illustrated in Figure~\ref{fig:teaser}. 
Our Technique, that we call Combined Walking in Place CWIP combines \textit{NW} with  \textit{WIP} with seamless transitions between the two using commodity depth sensors.
Though this technique people can access places in the VE that lie beyond the confines of the physical room available via \textit{WIP}. They can naturally switch to \textit{NW} when the point of interest in the VE is within the range of the physical space available.

Our navigation technique as typical of walking frees the hands to interacting with the VEs under the desired conditions. However, walking still poses the problem of how to inform the person on how to avoid going beyond the limits of the available space or hitting obstacles. This is because the person wearing an Head-Mounted Display (HMD) has no information about their position in the real world or on the obstacles in their path.
To solve these problems and thus allow safe locomotion in VR via CWIP we developed  a Warning System to assist people when  navigating VEs.
Our approach uses \textit{Visual Indicators} to inform the person of the existence and relative location of real-world obstacles to be avoided, such as the physical limits of the area available to move and other encumbrances, as depicted in Figure~\ref{fig:teaser}
Additionally, to increase people awareness of those obstacles we also implemented visual and audio signals to direct their attention to the augmented \textit{Visual Indicators}.
We named these techniques Augmented Virtual Reality (AVR) since they augment the VE with virtual objects superimposed to the visualization of the virtual world to convey information in a similar vein to Augmented Reality~\cite{azuma1997survey}.
Thus we call our combined approach CWIP-AVR resulting from combining CWIP navigation with AVR to implement collision Warnings.

The work described in the present paper aims to prove that CWIP-AVR enables people to walk immersive VEs within confined  and obstructed physical spaces. Furthermore, we conducted a user study to show that AVR provides for safe walking by keeping the user within limits and avoid obstacles in a less obtrusive manner than similar approaches.
Indeed informal comments reported by subjects, indicate that our approach presents a good trade-off between immersion and safety while providing safer locomotion.
\section{Related Work}

The interfaces used 
for locomotion in VEs can be divided into two main categories: direct and indirect~\cite{Mine1995}.
Direct control interfaces use the movement of body parts (such as head, arms, hands, torso, legs and feet, among others) ~\cite{Wilson:2016,Tregillus:2017} to control displacement in the VE.
Indirect interaction techniques use physical control devices to accomplish this task. Notable examples include keyboards, mouses, joysticks and gamepads. While these can be both effective, affordable, convenient and do not require much space, they do not provide an adequate sense of presence in the virtual world, outside of vehicular displacement (e.g. driving a car or piloting a spaceship among others), so we will will not delve into their particulars as we are interesting in the myriad applications where walking provides the most appropriate, elegant and apt metaphor to exploring the virtual reality.

Past research ~\cite{Usoh1999} shows that direct control interfaces convey the greatest sense of presence in the virtual world to humans. Among the techniques that implement direct control interfaces \textit{NW}~\cite{Ruddle2009, Heintz2009, zanbaka2005comparison, kuan2003constructivist, Janeh:2017,creepytracker,cigro,Metawidgets,Correia:2005} and \textit{WIP}~\cite{Slater_1995:TSI:210079.210084, Bruno2013, Bruno2017, Wendt2010, Feasel2008, Tregillus2016}
are some of the most relevant to the present research.
When using \textit{NW}, a person can walk both naturally and freely in a physical space and their position and orientation are replicated in the VE.
Indeed, \textit{WIP} allows the user to move forward in the VE by performing steps without moving expressively her or his position on the physical floor.
While this technique can convey a strong sense of presence in the VE this sense is not as strong as that afforded by \textit{NW} approaches~\cite{Usoh1999, Razzaque2002, yan2004new}. 

While \textit{NW} is the technique that best conveys a sense of presence in the virtual world, current implementations have requirements that the ordinary person can not meet. Example of these include the access to large open areas without obstacles that can be dedicated exclusively to navigating the VE.

The technique \textit{Redirected Walking}~\cite{Razzaque2002, Steinicke2008, Ngoc:2016, Grechkin:2016} is an example of an inadequate approach to \textit{NW} given the common limitations of most residential homes.
This approach consists in guiding a person by subtly changing the positioning of the VE at each step taken, so that he or she walks in cycles within the available area without noticing.
The main problem with this technique is the size of free space required ( 22m~\cite{Steinicke2009}) so that the human subject does not realize that he/she is not walking in a straight line.
While there are alternatives that do not necessitate such free space, those require the user to stop moving in the virtual world to reposition herself in the real world~\cite{williams2007exploring, Freitag2014}.

Using specialized apparatus such as the \textit{Omnidirectional Treadmill}~\cite{ruddle2011walking, souman2011cyberwalk, Souman2010} could provide a solution to this problem because these devices allow the user to walk freely in all directions without leaving the same place and some even can convey a sensation similar to \textit{NW}.
However, for reasons of cost and encumbrance this apparatus is not accessible to or affordable by everyday users on common living quarters.

As we can see from the above, current methods do not allow the average person to conveniently and safely navigate physically in VEs or to have the best possible experience. This is particularly true of home settings, where it is impractical to dedicate large unencumbered spaces to the fruition of VEs. This and a domestic limited budget makes
\textit{Omnidirectional Treadmills} largely inconvenient, while \textit{Redirected Walking} is either unfeasible or disruptive to the experience and \textit{NW} is limited to the existing physical space.
\textit{WIP} allows the physical navigation of VEs in a limited space but it affords less presence than \textit{NW}. What we need is therefore a technique that can combine the scalability of WIP and secondly to allow \textit{NW} within the limits of a confined environment, to match the characteristics of most consumers' homes. Finally, we need some way to allow people to safely avoid room walls and obstacles such as furniture, fixtures or pets while navigating immersive VEs, given that HMDs isolate the wearer from the outside environment.
\section{Our Approach}

As we seen previously, current navigation techniques and methods do not provide a natural or affordable counterpart to the commodity VR displays that have mushroomed in recent years. We present an approach that allows people to safely experience immersive VR by walking in the comfort of their homes. Our approach features a network of commodity depth sensors to detect the person's position and combines Walking in Place with Natural Walking in novel manners to achieve more flexible and powerful ways to support locomotion within domestic settings. Furthermore, our approach adopts Augmented Virtual Reality to make it possible to roam VEs safely within the confines of household dwellings.

\subsection{CWIP - Combined Walking in Place}
\label{sec:app_CWIP}
\begin{figure}[t]
    \centering
   \includegraphics[width=\columnwidth]{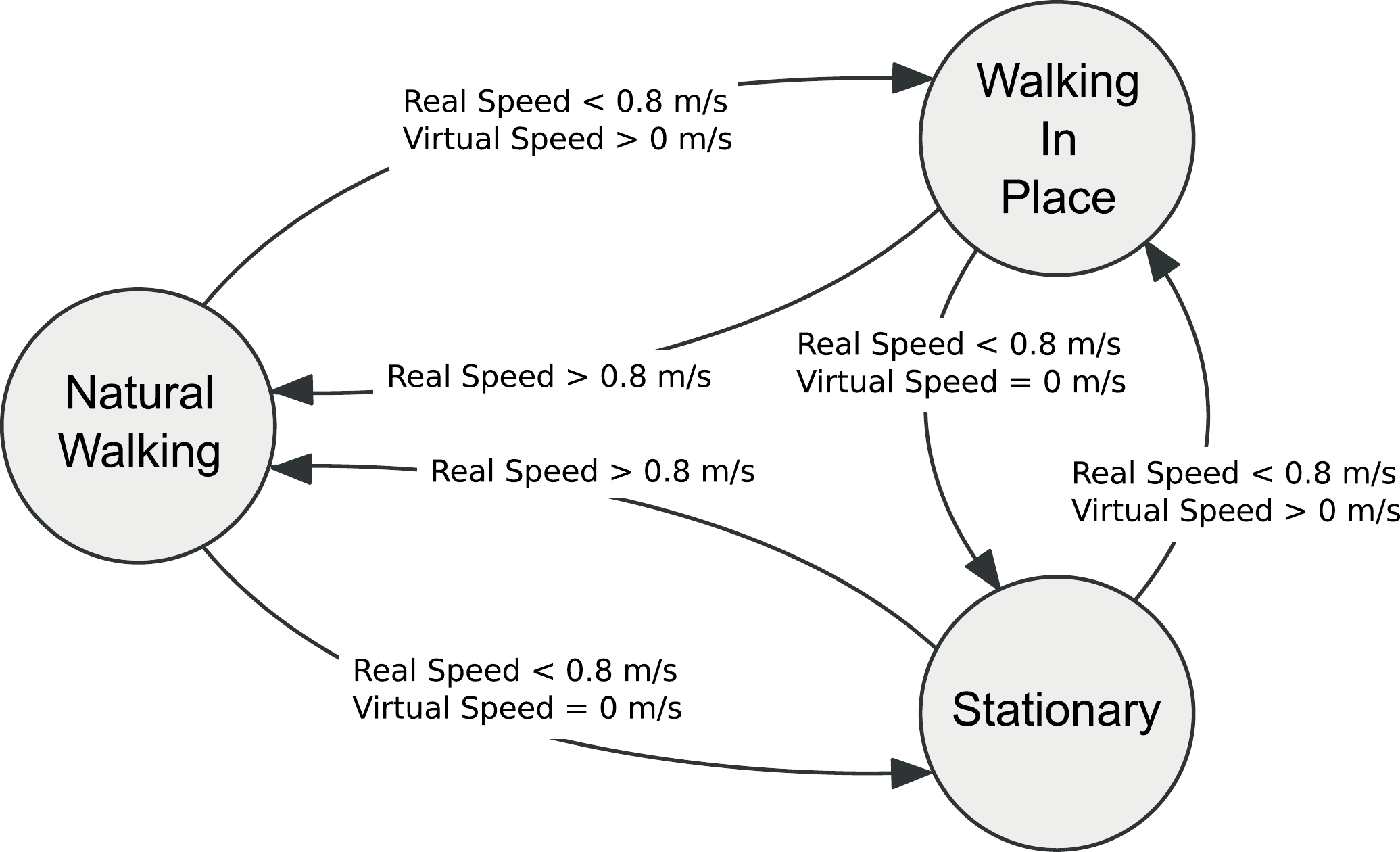}
   \caption{CWIP State Machine}
   \label{fig:ME_Impl}
\end{figure}

Our approach to navigation CWIP allows people to seamlessly and automatically alternate between two different locomotion techniques in the virtual world.
\textit{NW} is useful for within-reach displacements, especially circular movements around small objects of interest, in that it replicates the real world movement into the virtual world in that a person's steps map into their avatar in the VE.  \textit{WIP} is more suitable for far-reaching locomotion that cannot be readily accommodated within a small physical space, in that it induces virtual forward movement by simulating steps while marching in the same place in the real world.
To the best of our knowledge, our approach is novel in combining these two modalities in a natural manner. To this end, we determine which of the two techniques should be used at a given time according to the current state of a CWIP State Machine, according to three possible states: \textit{Stationary}, \textit{NW} or \textit{WIP}.
The state transitions are determined by two factors. First,
whether the person is in motion. Second, if not in motion, whether or not they are simulating steps.
If the system determines that the user is in motion (See Section~\ref{sec:CWIP}) the current state of the CWIP state machine, whose diagram is depicted in Figure~\ref{fig:ME_Impl}), is set to NW. As the name indicates, the movement of the avatar in the virtual world directly corresponds to their physical motion.

If the person is not walking and is not marching in place, the system is in the \textit{Stationary} state. In this state the avatar is at rest in virtual world, in a fashion similar to the behavior of the \textit{NW} state, in that the avatar follows the locomotion / immobility of the person.

If the user is not moving but is simulating steps, the system should be the \textit{WIP} state.
In this state, the movement of the avatar in the VE is determined by the locomotion technique \textit{WIP} (See Section~\ref{sec:CWIP}). 


\subsection{AVR - Augmented Virtual Reality}
\label{sec:app_AVR}

While CWIP allows for flexible locomotion, AVR is responsible for informing the person about the physical space limits and obstacles in their path,
during their navigation in the virtual world.
To accomplish this, we are augmenting the Virtual World representation using three distinct elements: \textit{Visual Indicators}, \textit{Visual Markers} and \textit{Sound Alerts}.
The \textit{Visual Indicators} adds synthetic Objects to the Virtual World, that behave independently of it and have a distinct visual appearance. They serve to signal the location of hazards and obstacles to progression in the real world.
These indicators can be divided into two Groups: Limits are represented by translucent planes that depict the limits of the space available for physical in front that the person; Obstacles, rendered as solid parallelepipeds, indicate the location of obstacles in the Real world, such as chairs, tables, beds and other pieces of furniture, that lie in the path of the user. 

\begin{figure}[!tp]
   \centering
    \includegraphics[width=0.45\textwidth]{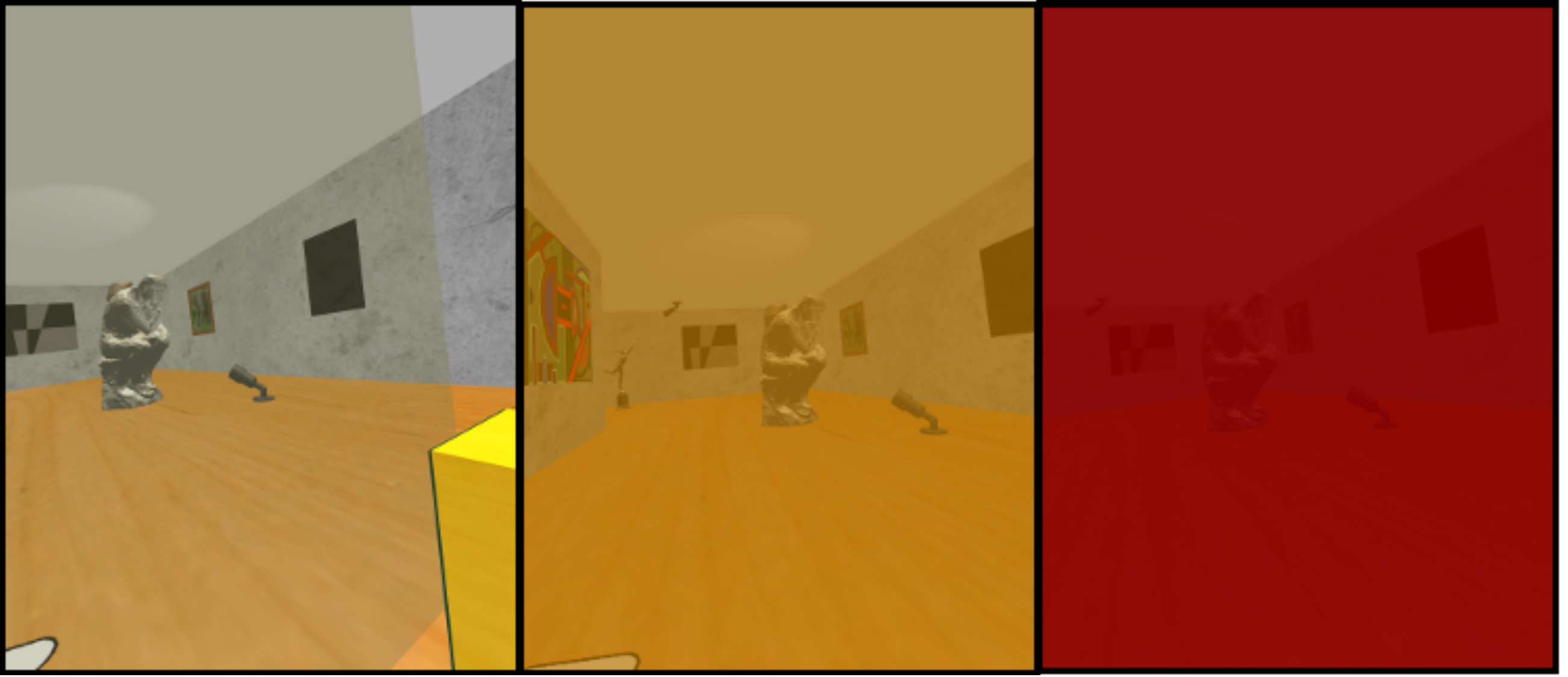}
    \caption{Rendering a Visual Indicator of a limit in different zones (from left to right: Pre-Warning, Warning, Danger)}
	\label{fig:Avr_Limites}
\end{figure}

An \textit{Indicator}'s, color and transparency, depends on its distance to the person. Distinct areas where the Indicator has a similar behavior are called \textit{Zones} according to proximity. There are four of these: \textit{Normal Zone}, \textit{Pre-Warning Zone}, \textit{Warning Zone} and \textit{Danger Zone}, as depicted in Figure~\ref{fig:Avr_Limites}.

These Zones can be divided into two subgroups: Constant, where the aspect of the indicator is always the same regardless of the exact distance to the user, and Dynamic, where the aspect linearly depends  on the distance the user is between an upper and lower limit specific to that \textit{Zone}

\begin{figure}[!bp]
   \centering
    \includegraphics[width=0.45\textwidth]{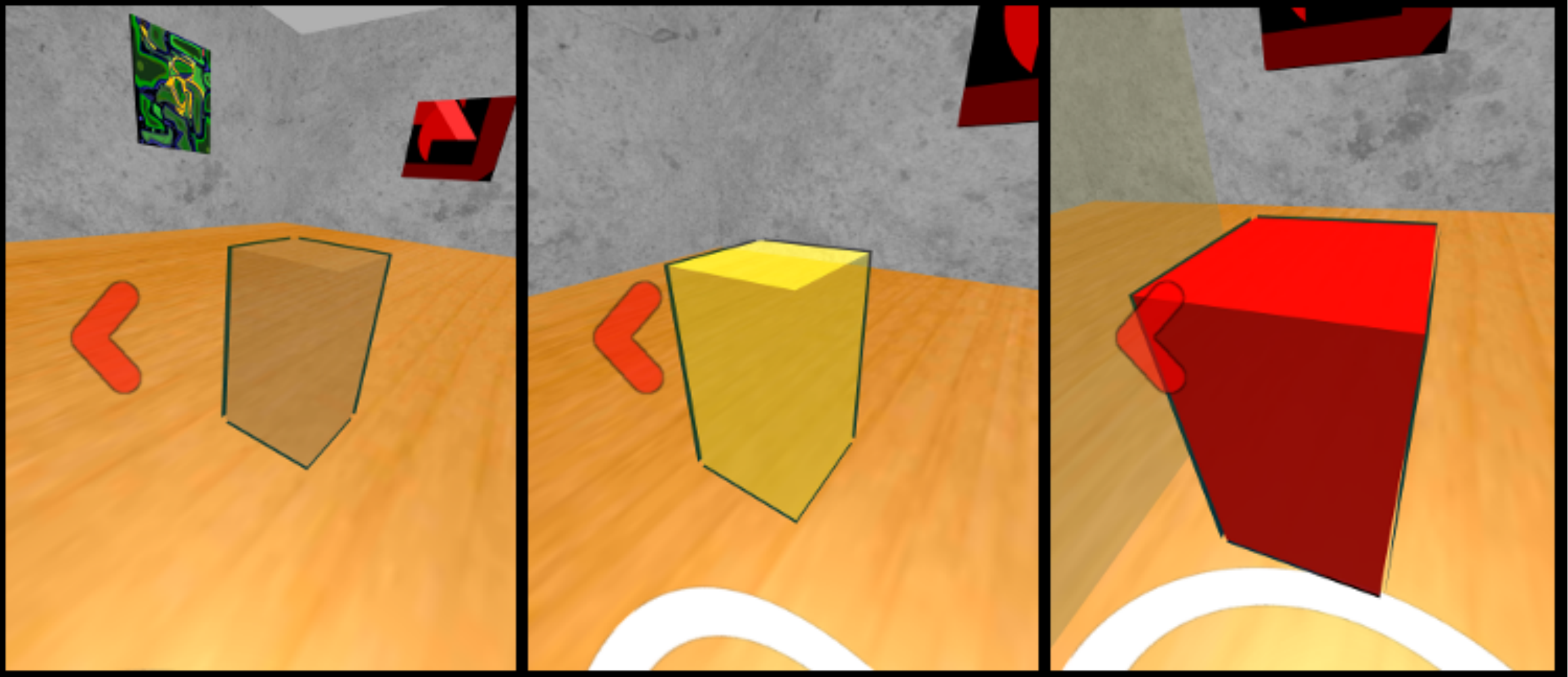}
    \caption{Rendering a visual indicator of obstacle in different zones. The red Visual Marker indicates the presence of another obstacle to the left of the user (from left to right: Pre-Warning, Warning, Danger)}
	\label{fig:Avr_Obstaculos}
\end{figure}
There are distinct behaviours for each Zone.
Within the \textit{Normal Zone} the \textit{Visual Indicators} are invisible.
Within the \textit{Pre-Warning Zone} the \textit{Visual Indicator} behaves Dynamically from being transparent and white in the upper limit to linearly progressing to yellow and Semitransparent once its lower limit is reached.
Inside the \textit{Warning Zone} the \textit{Visual Indicator} behaves Dynamically from being yellow and semitransparent at the upper limit, and linearly changing  until it becomes red and almost opaque in the lower limit.

When in the \textit{Danger Zone} the \textit{Visual Indicator} features a Constant behavior being rendered as red and almost opaque.

\textit{Visual Markers} inform the person of real-world hazards. However, if the user not looking in their direction there is no way for \textit{Visual Indicators} to inform on the state of the real world.
To approach this problem, we included two types of \textit{Warning Signs} that direct the user's attention to the Visual Indicators.
\textit{Visual Signs} (see the Figure~\ref{fig:Avr_Obstaculos}), are red arrows depicted on the side of the FoV, to indicate the direction of obstacles that may be visible nearby in the virtual world but out of the person's sight.
\textit{Sound Signals} are audible alarms  that alert the user that they about to enter the \textit{Danger Zone} of an obstacle outside their field of view.
These alarms can stop in one of two situations.  Either the person leaves the \textit{Danger Zone} of the hazard or they look directly at the corresponding \textit{Visual Indicator}, thus showing that they are aware of the existence of the obstacle.
\section{Implementation}

In this section we describe our proof of concept prototype, using an array of depth cameras. Occlusion free body tracking is done via an in house open-source toolkit that is described elsewhere as an accepted publication at an ACM conference~\cite{creepytracker}. 

\subsection{Apparatus} 
\label{section:SetUp}

As said above, our system uses an array of evenly spaced depth sensors to track a person´s (or people) position and body joints.
To function correctly our CWIP-AVR approach requires a steady stream of information about the person including information on their body joint locations and motion within the tracking area.
To meet this requirement, it is necessary that the sensors are arranged so as to ensure that in the entire interaction area, at least one sensor is able to capture the points of interest (aka skeleton) of the user, notably the \textit{Knees}, \textit{Chest} and \textit{Head}.
Our prototype implementation uses five \textit{MS Kinect One} cameras, placed in the vertices of an irregular pentagon around the tracking area.
The information captured by these sensors is processed by a Tracker program using the above-mentioned toolkit~\cite{creepytracker} to produce information about the points of interest, selected from the most reliable sensor at any given time.

We chose Kinect because of their affordable price and most importantly because they do not require special markers, as is the case of competing setups such as \textit{Optitrack} which require optical markers and come at a higher price point. We did not use controllers, such as the \textit{HTC Vive}, since the existing implementations require active markers and they did not allow tracking enough points of interest to implement our approach. Using commodity depth cameras, however, comes at the cost of less precision when compared to optical tracking systems, and by nature the data are more affected by noise. However, our approach does not require highly precise data and the toolkit provides a good compromise between smooth temporal filtering and lag. 
As a result, our setup can reliably track up to four people over an area voluntarily restricted to 3mx3m, where all the relevant body joints can be reliably tracked by at least one the five cameras regardless of obstacles, and where a person has enough free space to walk three to four steps in a straight line, modulo the furniture.

As for immersive visualization, we used a \textit{Gear Vr} apparatus to visualize the virtual world because it does not need power cables or an external desktop computer, unlike \textit{Oculus Rift} or \textit{HTC Vive}. This avoids many potential problems with \textit{NW}, such the person tripping on the cable ot it becoming entangled on the obstacles. The depth cameras operate over a local network and of course obstacle information has to be transmitted over wi-fi. The setup features also less computational power and less autonomy than a desktop setup, due to its dependence on a smartphone battery. Due to these limitations, we relied on precomputed scans of the real environment, instead of using real-time cellphone-based reconstructions of the physical environment, which would have killed real-time performance. While our precomputed setup may lack a bit of flexibility, this is not central to or detrimental of our key contributions. In the future, we plan to add real-time reconstruction to our toolkit, in a setup reminiscent of RoomAlive~\cite{Jones:2014:RME:2642918.2647383} wrt capturing the geometry of the room, which will then be parsed for obstacles, such as fixtures, furniture, other people or pets. However, the current prototype is well-suited to illustrate the virtues of our approach to locomotion in VR. This scenario also fits well with current visions of future domestic VR setups.

The prototype was implemented on top of the \textit{Unity3D} graphical engine because this framework provides a ready-made toolbox to support virtual reality projects for common setups such as the \textit{Oculus Rift} or \textit{Gear Vr} apparatuses.


\subsection{CWIP - Combined Walking in Place}
\label{sec:CWIP}


As we have seen above, CWIP relies on a state machine fed by sensor information to determine whether the person is moving and whether they are either marching in place or naturally moving. We determine heuristically the rate of speed by measuring the variation of the horizontal position of the person's \textit{Chest}, as communicated by the depth sensors.
We chose the \textit{Chest} because it is the most reliably captured joint by the depth sensors and because it is the  most stable  point of the body during \textit{WIP}. We discussed at length how to determine the threshold which causes the state machine to move to NW state. We chose this value as an average step per second, a distance of  ($0,70-0,75 m$) for an average-height person walking at a leisurely place. Due to the inaccuracy and noise of depth sensor data and the tracker used we found it necessary to introduce an experimentally determined safety margin to minimize situations of \textit{WIP} being recognized as \textit{NW} and vice-versa. We found out that we achieved a recognition rate of 90\% to 93\% which seemed comfortable to subjects during experimental evaluation.

To compute the speed limit, we observed over a dozen people walking under experimental conditions (see the evaluation section below) and used a Boxplot~\cite{chambers:1983:GMDA, mcgill1978variations} to identify the outliers caused by the imprecision of the system and by the participants inadvertently wandering  forward when they were instructed to exercise the \textit{WIP} movement. 

The upper speed threshold thus obtained ($0,78m/s$) represents $94.60\%$ of the accumulated percentage of the speed frequency distribution. Thus, for added stability we chose a threshold of $V_t = 0,80m/s$ for the CWIP state machine to trigger the transition between states.



Our technique builds on the motion state machine \textit{WIP} developed by Bruno et al.~\cite{Bruno2017} who used a singe depth camera to detect the steps taken while stepping in place in front of a TV screen. Based  on  the pace and maximum height of the knee joints during the steps the \textit{WIP} state machine determines the corresponding virtual speed. This is then used to compute the forward movement of the person's avatar in the virtual world.

We have extended this approach further to allow people to alternate between \textit{WIP} and \textit{NW} once the above threshold velocity $V_t$ is crossed by the person who is navigating the virtual world. We acquire the height of the knee joints via the same \textit{Tracker} data used to estimate the position of the chest to compute the virtual speed when \textit{WIP}.

To deal with the imprecision and noise problems mentioned above, the captured instantaneous position values of the user's body joints are smoothed using a \textit{Kalman}~\cite{Kalman1960} filter.

\subsection{AVR - Augmented Virtual Reality}
\label{sec:AVR}

For the above approach to work, we need expeditious ways to make people aware of real world boundaries and obstacles when they are immersed in the Virtual World. As we described earlier, AVR conveys this information as unobtrusively as possible by encoding proximity and line-of-sight information and synthetic virtual objects and markers. Proximity relies on several parameters that relate to body proportions and reach of subjects when they walk around in a closed space. 
The parametric distances that define the size of each zone were empirically determined in relation to the size of a step (about $70cm$) taken by an average person.

Due to the same noise and flicker problems mentioned in the section~\ref{sec:CWIP} we added a safety margin of $20cm$ empirically validated.

\begin{figure}[!tp] 
    \centering 
    \begin{minipage}[b]{.375\textwidth}
        \includegraphics
         [width=\textwidth]
        {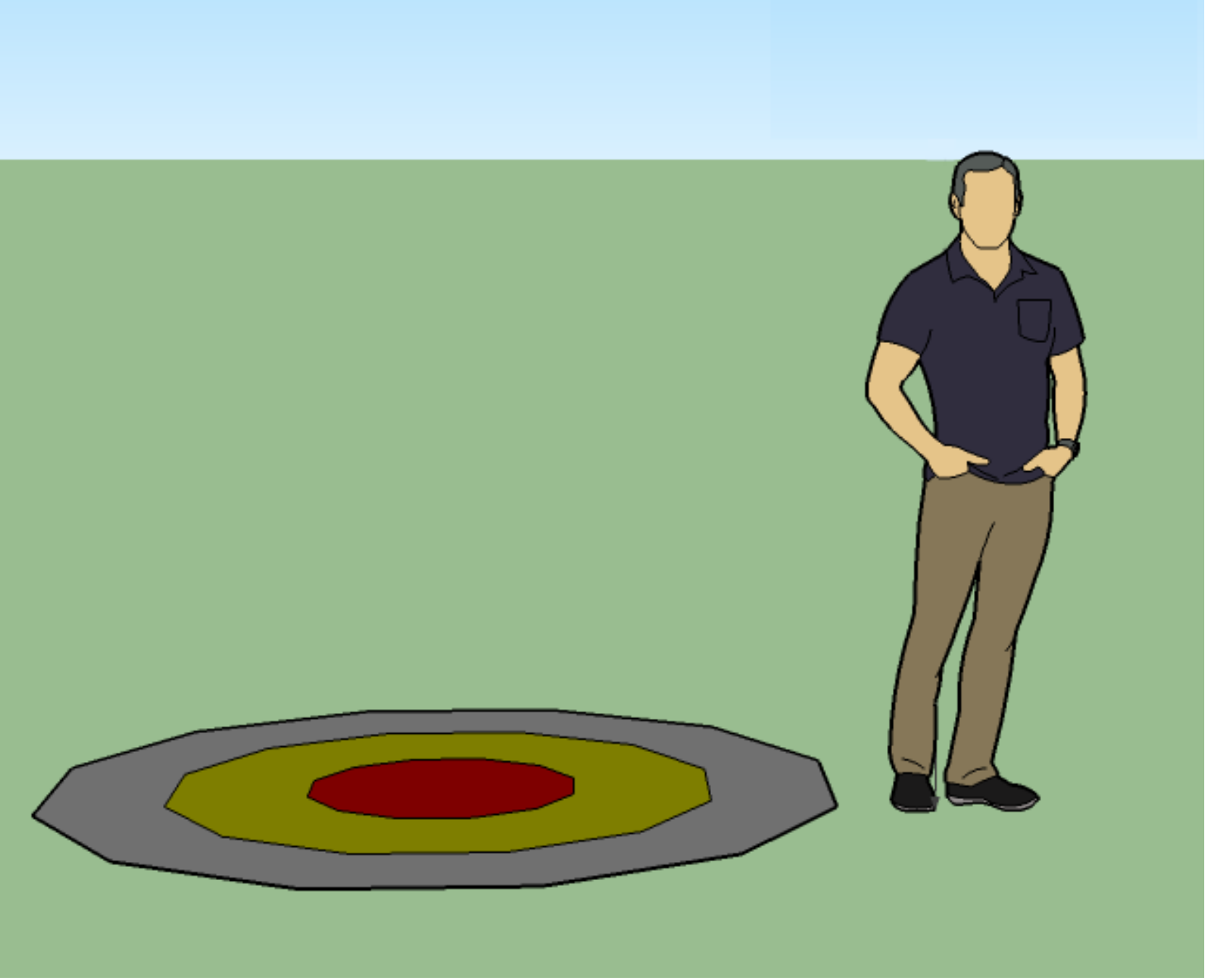}
        \caption{Normal Zone}
    	\label{fig:ZonaNormal}
    \end{minipage}
    \hfill
    
    \begin{minipage}[t]{.375\textwidth}
        \includegraphics
         [width=\textwidth]
        {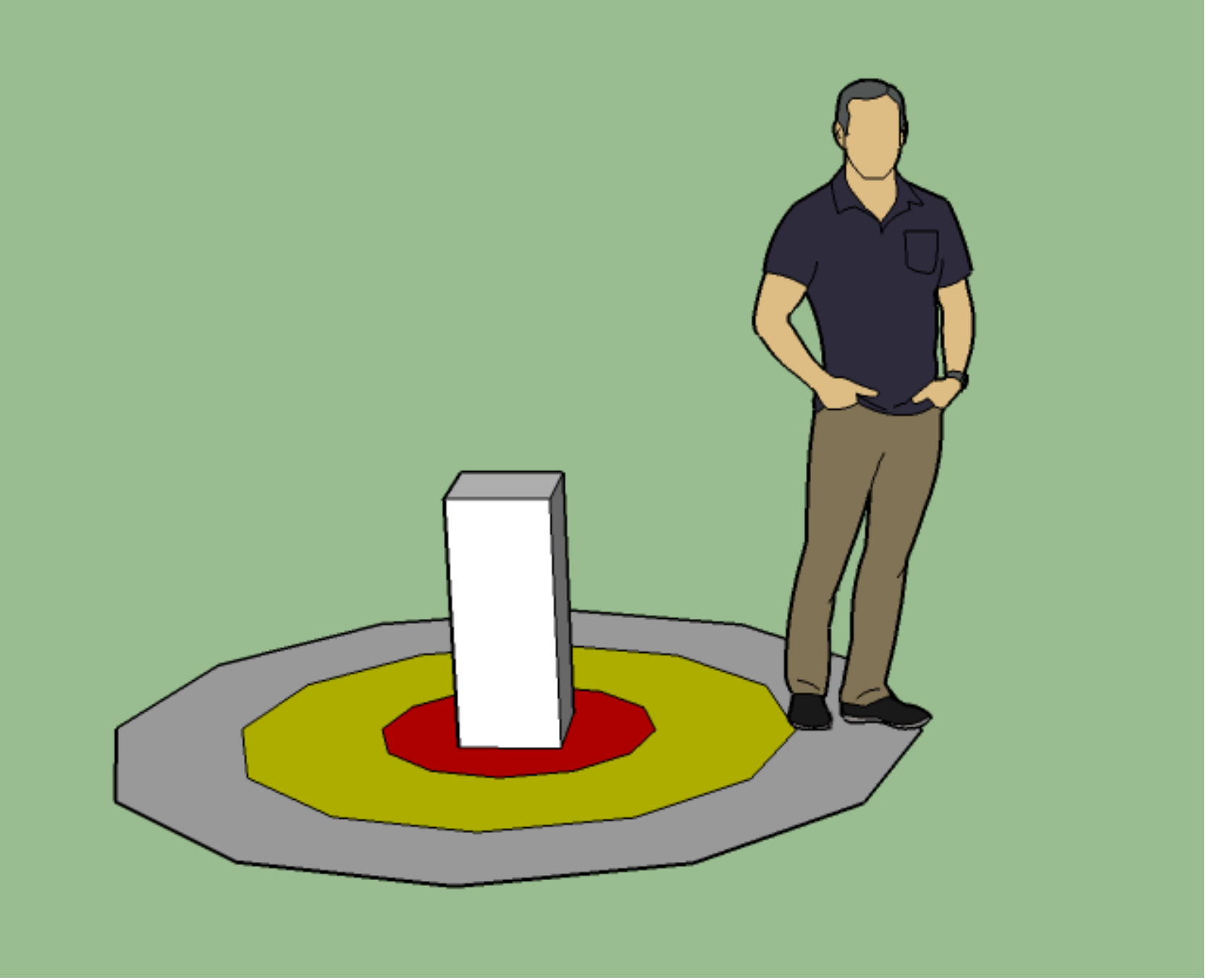}
        \caption{Pre-Warning Zone}
        \label{fig:ZonaPreAviso}
    \end{minipage}
    \hfill
    \begin{minipage}[h]{.375\textwidth}
        \includegraphics[width=\textwidth]
        {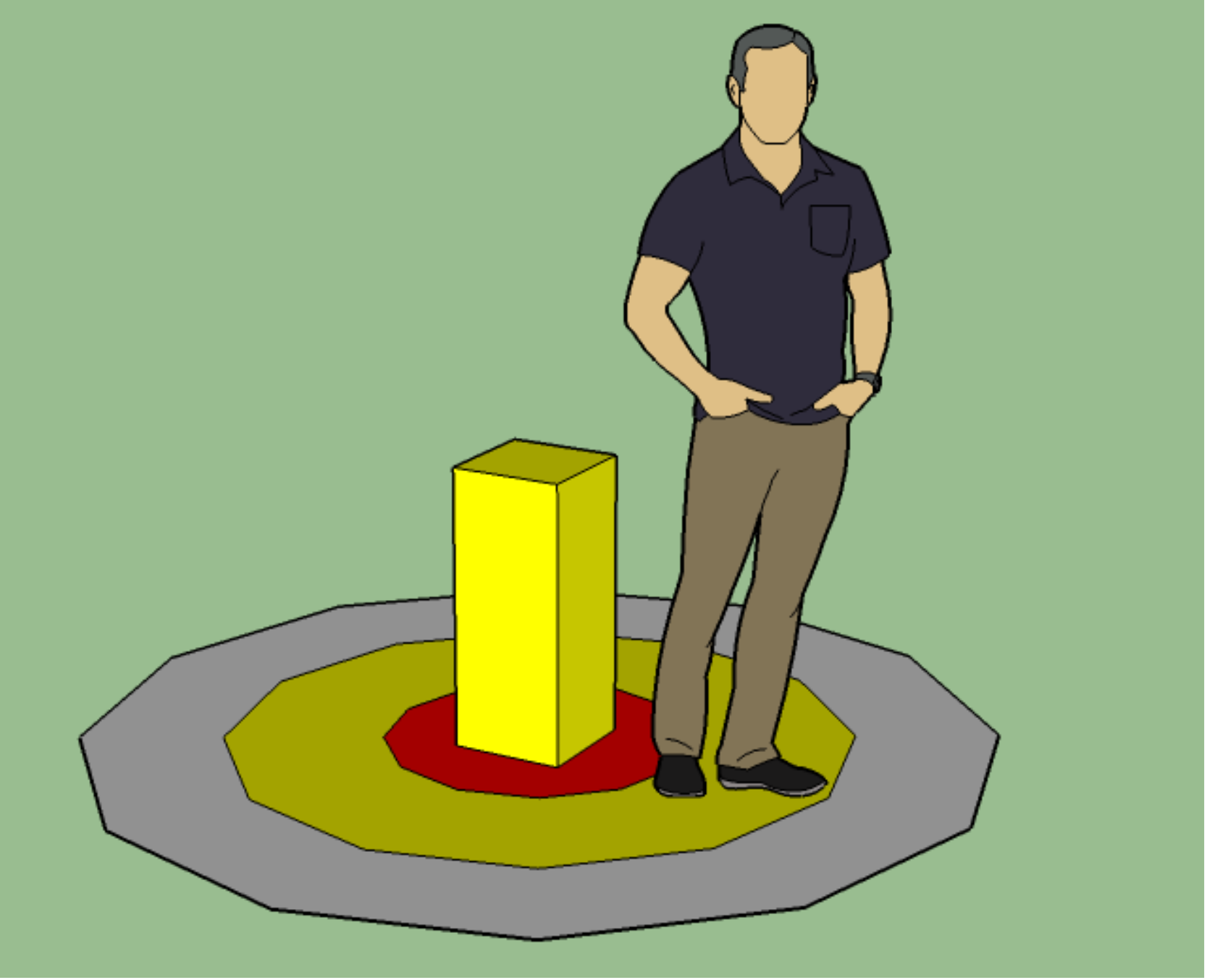}
     \caption{Warning Zone}
        \label{fig:ZonaAviso}
    \end{minipage}
    \hfill
    \begin{minipage}[b]{.375\textwidth}
        \includegraphics[width=\textwidth]{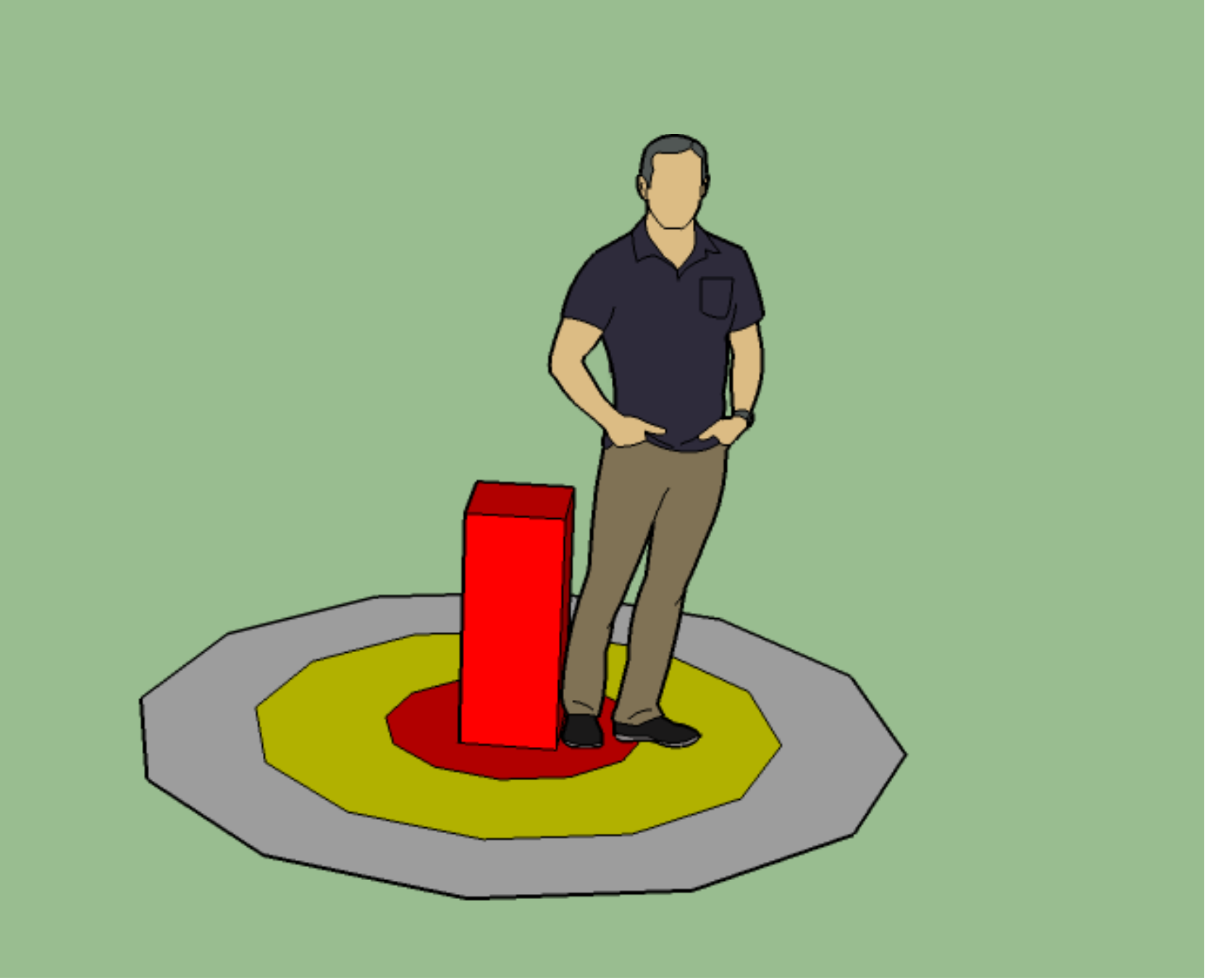}
        \caption{Danger Zone}
        \label{fig:ZonaPerigo}
    \end{minipage}   
\end{figure}

The upper limit of the \textit{Danger Zone} is then based on distance covered by a half-step. This distance, including the safety margin is equal to $0.40m$, as depicted in Figure~\ref{fig:ZonaPerigo}.

We defined the upper limit of the \textit{Warning Zone} based on the distance covered by an average step as $0.80m$, as shown in Figure~\ref{fig:ZonaAviso}.

We defined the \textit{Pre-Warning Zone} using the distance corresponding to a step and a half or $1.20m$, as exemplified in Figure~\ref{fig:ZonaPreAviso}. Above this distance the user is in the \textit{Normal Zone}, as illustrated by Figure~\ref{fig:ZonaNormal}.




\section{Evaluation}
\label{sec:eval}

\begin{figure}[htp]
   \centering
    \includegraphics[width=0.45\textwidth]{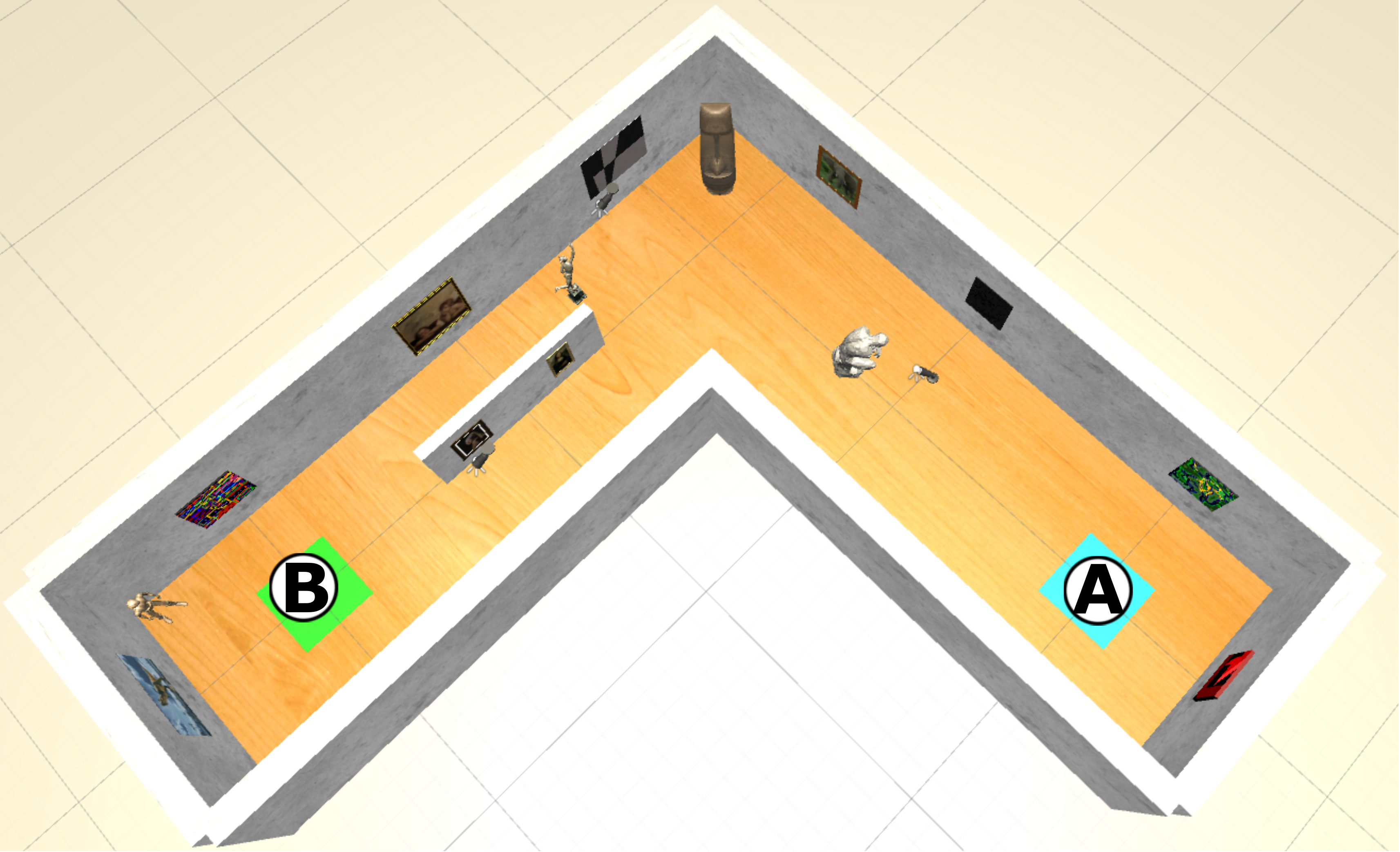}
    \caption{Virtual environment used in the test task. 
    (Seen from above),  A - Starting point of the Test , B - End point of the Test}
	\label{fig:MuseuAB}
\end{figure}

To validate the CWIP-AVR approach and evaluate its performance we carried out a user study featuring 20 participants (16 men and four women) with ages between 17 and 58, most of them in the 25-36 year age group ($55 \%$). The user test was carried out on our lab due to logistics although not all the subjects were graduate students.

\begin{figure}[b]
   \centering
    \includegraphics[width=0.45\textwidth]{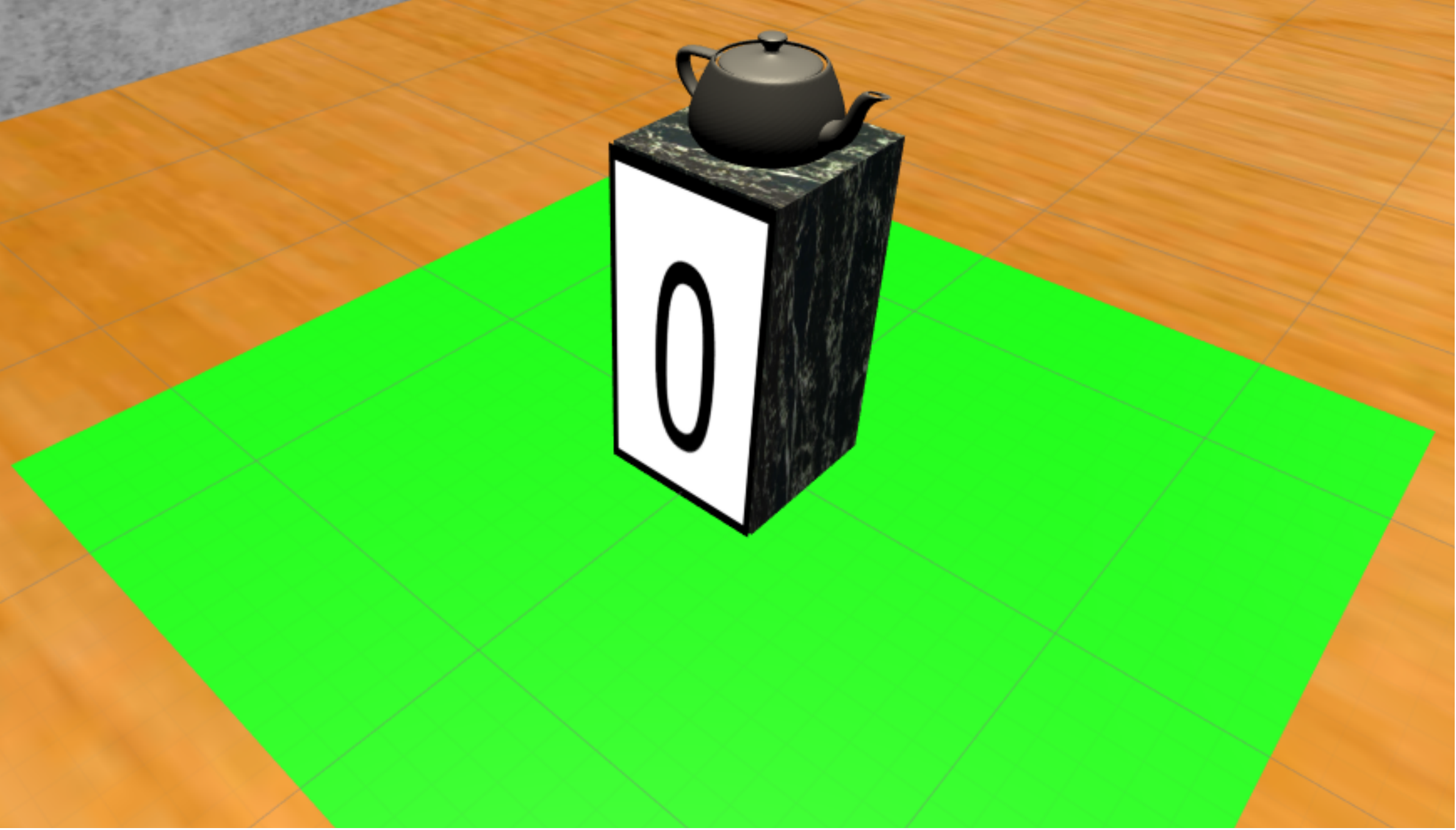}
    \caption{Example Task Goal}
	\label{fig:TaskGoal}
\end{figure}

The study consisted a series of three identical tests (with an extra free form initial training session) in which each participant had to accomplish a task in a virtual world using the proposed approach, using ours and a baseline approach, chosen at random.
The task consisted of navigating a VE (see Figure~\ref{fig:MuseuAB}),  featuring a virtual art gallery twenty meters long up to a 90 degree curve followed by another twenty-meter long gallery leading to a podium with an object on top. 
In this podium there was a random number that the participant had to say aloud to finish the test (See Figure~\ref{fig:TaskGoal}).


\begin{figure}[!bt] 
    \centering 
    \begin{minipage}[b]{.375\textwidth}
        \includegraphics
         [width=\textwidth]
        {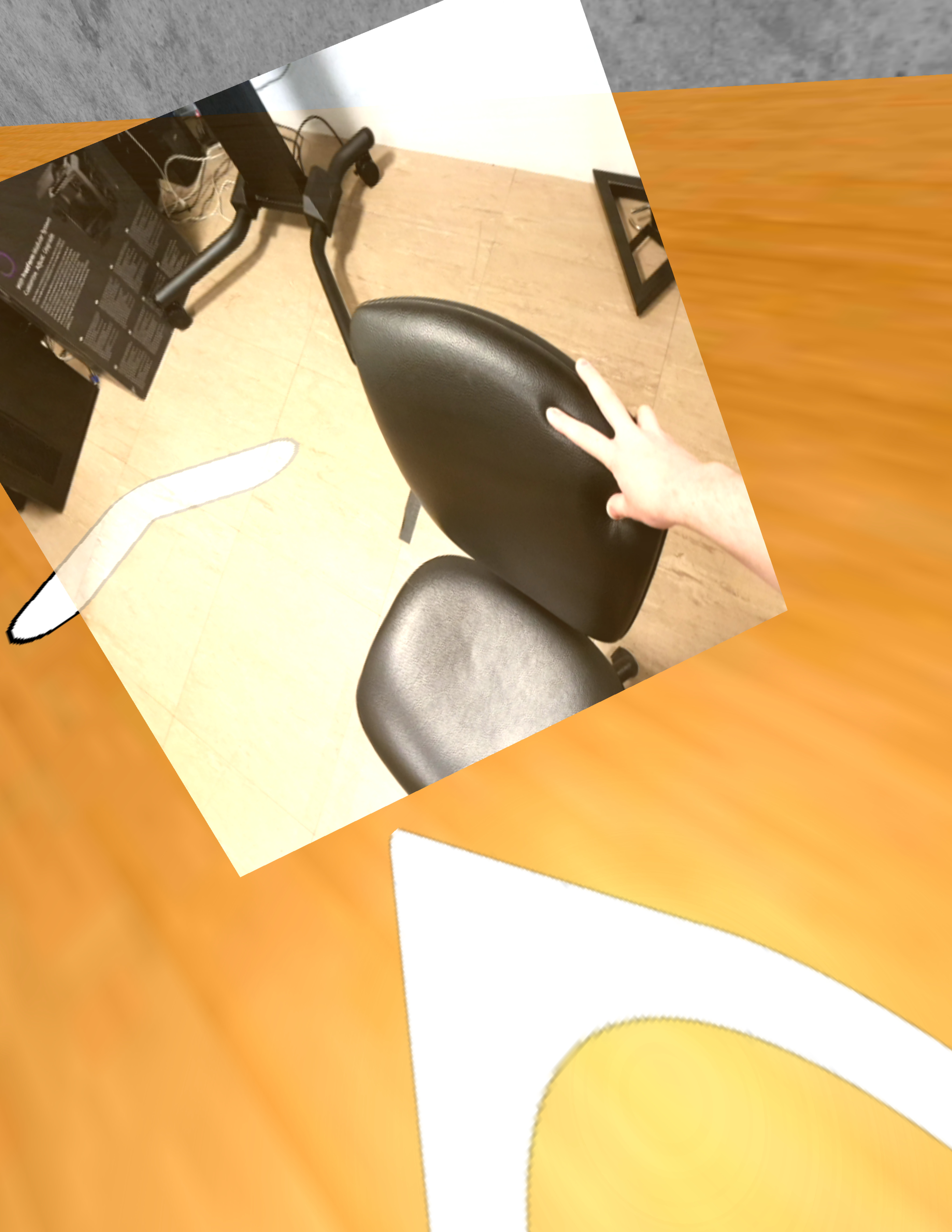}
    \end{minipage}
     \hfill    
    \begin{minipage}[b]{.375\textwidth}
        \includegraphics
         [width=\textwidth]
        {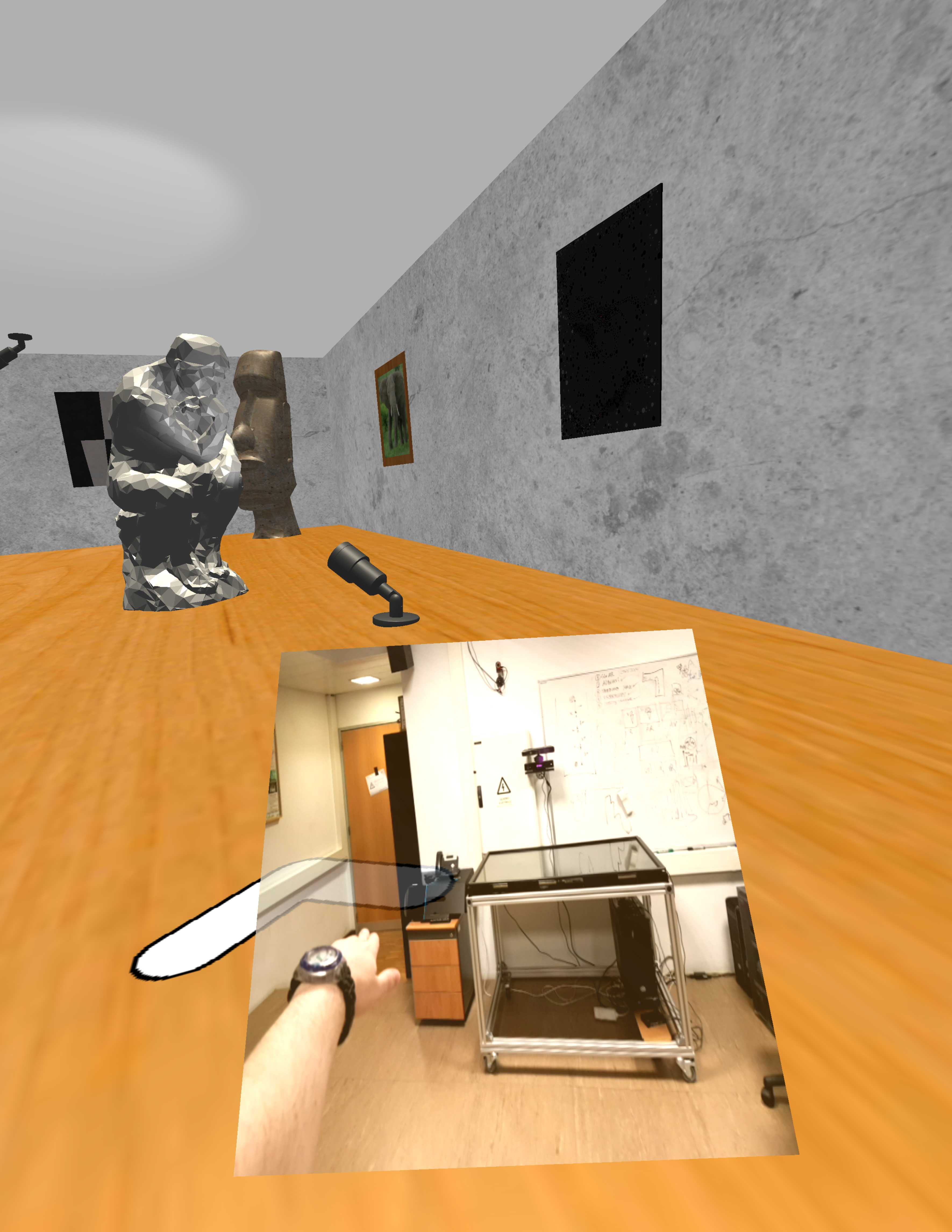}
    \end{minipage}
    \caption{Baseline Implementation Examples}
     \label{fig:Baseline_Exs}
\end{figure}

All participants were able to complete the proposed challenges, while alternating at will between WIP and NW which demonstrates the validity of our approach. 
To evaluate the performance of our proposed Warning System AVR, series of identical tests were performed using the proposed CWIP navigation approach against the baseline System, called \textit{Passthrough} (see Figure~\ref{fig:Baseline_Exs}). We chose pass-through video since it is readily available on our development framework and it allowed a direct comparison to our technique, since the goals are the same (allowing people to safely walk in immersive VEs).
This System consists of a translucent plane at ground level where the image captured by the \textit{Gear Vr} camera is mapped. This system allows participants  to check for themselves whether they are in risk of colliding with real world obstacles during  navigation.
This approach is an adaptation of \textit{Steam Vr's} \textit{Chaperone} system to \textit{Gear Vr}.
The \textit{Chaperone} system transmits the image captured by the \textit{Steam Vr} camera to a plane next to the controller. As \textit{Gear Vr} has no equivalent controllers we placed  the plane at a fixed position below the user's waistline, as a compromise between being readily accessible and not interfering too much with navigating the virtual world.

\subsection{Evaluation Metrics}
\label{sec:evalMetrics}

The two approaches were evaluated according to user preferences and different task performance metrics.

To assess user preferences, we had the participants answer a questionnaire after completing each series of tests for either approach.
The questionnaires were presented using a Likert scale with six values. This served to determine the unambiguous participants' opinion on the effect each technique had on the perception of the real-world vs the Virtual World. 
The questionnaire consisted of five questions related to the ease of identifying certain elements of the task (participant's position in the real and virtual world, the position of the walls and the obstacles (chairs), and the ease of achieving the task goal). We included also two questions about the effects of using the system (tiredness and discomfort).

We assessed Task Performance according to the values recorded during the execution of each test using three relevant metrics: task time, the number of exits from the limits of the tracking area (note that we set this a little behind the physical lab walls for safety), and the number of  obstacles hit.

We measured the time that each subject took to complete the task.
To reduce the effect of habituation to the approach in the result, tests were randomized and the value used in the analysis was the mean value of the three tests performed.

We also noted the number of boundary crossings (wrt tracking area). This was the number of times the participant during the execution of the task in the virtual world  inadvertently left the tracking area 
While this does not account for actual wall collisions in that situation one time off the limits could mean the end of the experience.
To take this into account, we decomposed this metric into two: the total number of tests where at least once the participant went beyond the limits and the number of times the participant left and re-entered the limits of the tracking area.

Finally, we also counted Obstacles hit, every time that a participant during the experiment hit one obstacle, in this case chairs, in the real world.
Because the obstacles used were padded light chairs, hitting them did not seemingly bother the participant too much  (except for the element of surprise) so is possible to hit a chair and continue without stopping the experience which might not be true of collisions with heavier non-padded furniture or pets or even other people.
We also decomposed this metric into two: The total number of tests where a crash occurred and the total number of crashes that occurred in all tests.

In addition to these metrics, we also recorded all instantaneous speed data during the training runs, to 
empirically compute the $V_t$ threshold value used by CWIP to determine whether the user naturally moving vs standing or walking in place.
To this end, we analysed the users' speed values during a part of the test run where they were told to use \textit{WIP} (In the figure~\ref{fig:MuseuAB}, between point A and the bend).

\subsection{Results}
\label{sec:resul}

All participants were able to learn and use the CWIP locomotion technique and all were able to complete every test. This demonstrates the feasibility and appropriateness of our contribution. 

To analyze the questionnaires, we applied the \textit{Wilcoxon Signed-Rank}~\cite{wilcoxon1945individual} test to determine whether the differences in the results were statistically significant. From this analysis, we conclude that only the first question ("During the Task was it easy to understand what was my position in the Real World") presents a statistically significant result ($p-value=0.044$).
According to the user responses, we can conclude that the \textit{Passthrough} approach was preferred to the AVR as to the ability of subjects to perceive their position in the real world, which does not come out as surprising given the explicit video information they were given.
The values of the remaining responses do not present statistically significant differences for either approach.
We can only surmise that participants considered that in both approaches it was easy for them to determine the location of the boundaries and obstacles of the  real world, while maintaining the perception of their position in the virtual world and perform the intended task.

When evaluating task performance for each approach, the value of the \textit{task time} metric, used to determine the efficiency of each technique, is the average duration of the three tests that participants performed for each method. Our results indicate $102.006s$ (Std dev (SD) $44.61$) seconds for the AVR approach and $70.290s$ (SD $32.77$) seconds for the \textit{Passthrough} approach). The large SD values illustrate the wide variability in results, which is symptomatic of the widely different skills, familiarity and backgrounds in our subject population.

We applied a \textit{Shapiro-Wilk}~\cite{shapiro1965analysis, razali2011power} normality test to these two samples. The statistical results ($p-value=0.039$ for the approach AVR and $p-value=0.004$ for the \textit{Passthrough} approach) show that neither sample is normally distributed.

We then performed a \textit{Wilcoxon Signed-Rank} statistic test. The test result ($Z=-2.987$, $p-value=0.003$) shows that there is a statistically significant difference between the times measured on the two approaches. It means that the \textit{Passthrough} approach takes less time than the AVR approach to perform the task.

In comparing both techniques for failures (AVR and \textit{Passthrough}) we observed that participants collided  with real-world obstacles in four tests. However the total number of hits observed across all tests were four times with AVR and five times when using \textit{Passthrough}. This means that both approaches allow users to achieve comparable performance when avoiding obstacles. However, during informal post-test questionnaires, many participants complained about feeling less present when using video Passthrough than when experiencing AVR.

The total number of boundary crossings however, stakes three occurrences with the AVR condition against 13 incidents under the \textit{Passthrough} condition. Moreover, the total number of times participants left the tracking area across all tests total six times with the AVR approach against 46 times under the \textit{Passthrough} condition.

Because the AVR method was responsible for a considerably smaller number of exits of the designated area, relative to the number of occurrences and the total number of tests, it can be concluded that the AVR approach works best in keeping participants safe and within the interaction area. 



\section{Conclusions}
\label{sec:concl}


As we can observe from test results, all participants were able to master the Combined Walking in Place technique with little effort. This establishes our approach as an effective way to navigate virtual environments,using commodity depth sensors.
The results of the questionnaires and the \textit{task time} metric indicate that the \textit{Passthrough} technique is more efficient and the best to communicate the position of the person in the real world. however, this was judged by participants as breaking the sense of presence in the VE. One participant likened this approach to "PIP TV without the joystick".

Although the participants had gotten the impression that with the \textit{Passthrough} approach they had a better perception of their position in the real world, using the \textit{Passthrough} approach resulted in a considerably greater number of exits from the designated area. Furthermore, the results of collisions with the boundaries of the tracking area consolidate the perception of \textit{Passthrough} as the unsafer method. Indeed, each exit could have resulted in the participant hitting the wall if there were no safety distance between the limits of the tracking area and the physical lab walls.
Had the test been more stringent, that is if leaving the limits stopped the test, this would have resulted in 13 failed tests with the \textit{Passthrough} and 3 with AVR conditions respectively.

We have identified a trade-off between the better safety of the AVR approach versus the greater efficiency of the \textit{Passthrough} approach.
However, users might have difficulty taking advantage of the greater efficiency the \textit{Passthrough} approach because of the danger of crashing into the wall of the room thus interrupting the experience in the VE. This makes AVR the better approach. 

We have proposed CWIP-AVR, a new approach to navigating VEs using consumer-grade hardware in domestic settings by unsophisticated users.
From our user test results, we conclude that our technique can be applied to navigate safely in confined and obstructed physical spaces without significantly limiting the user experience and presence in the virtual world.

We believe that there is still much work to be done, regarding navigation and interaction with VEs in domestic and everyday scenarios, if the new technologies are to take hold in consumer domains. Indeed, our toolkit can and should be extended to include real-time acquisition and reconstruction of cluttered room settings. Also, further user testing and enhanced state machines can improve the user experience with AVR. We also plan to experiment with different AVR techniques to find out more effective reality awareness techniques. With the ever expanding offers for immersive and affordable HMD apparatus, VR and AR will further expand their reach and implantation as new media in the years to come, if research continues to further develop novel, effective and natural ways to interact with the new content.

\balance{}

\bibliographystyle{SIGCHI-Reference-Format}
\bibliography{main}

\end{document}